\documentclass[aps,prl,reprint,superscriptaddress,amsmath,amssymb]{revtex4-1}

\pdfoutput=1

\usepackage{graphicx}
\usepackage[pdftex,breaklinks=true,bookmarksopen=true,bookmarksopenlevel=3,bookmarksnumbered=true,colorlinks=true,urlcolor= magenta,citecolor=blue,linkcolor=blue]{hyperref}

\begin{document}

\title{Quasi-monoenergetic electron beams production in a sharp density transition }

\author{S. Fourmaux}
\email{fourmaux@emt.inrs.ca}
\affiliation{INRS-EMT, Universit\'e du Qu\'ebec, 1650 Lionel Boulet, Varennes J3X 1S2, Qu\'ebec, Canada}
\author{K. Ta Phuoc}
\affiliation{Laboratoire d'Optique Appliqu\'ee, ENSTA ParisTech - CNRS UMR7639 - \'Ecole Polytechnique ParisTech, Chemin de la Huni\`ere, 91761 Palaiseau, France}
\author{P. Lassonde}
\affiliation{INRS-EMT, Universit\'e du Qu\'ebec, 1650 Lionel Boulet, Varennes J3X 1S2, Qu\'ebec, Canada}
\author{S. Corde}
\affiliation{Laboratoire d'Optique Appliqu\'ee, ENSTA ParisTech - CNRS UMR7639 - \'Ecole Polytechnique ParisTech, Chemin de la Huni\`ere, 91761 Palaiseau, France}
%\author{S. Payeur}
%\author{S. Gnedyuk}
\author{G. Lebrun}
\affiliation{INRS-EMT, Universit\'e du Qu\'ebec, 1650 Lionel Boulet, Varennes J3X 1S2, Qu\'ebec, Canada}
\author{V. Malka}
%\author{S. Sebban}
\author{A. Rousse}
\affiliation{Laboratoire d'Optique Appliqu\'ee, ENSTA ParisTech - CNRS UMR7639 - \'Ecole Polytechnique ParisTech, Chemin de la Huni\`ere, 91761 Palaiseau, France}
\author{J. C. Kieffer}
\affiliation{INRS-EMT, Universit\'e du Qu\'ebec, 1650 Lionel Boulet, Varennes J3X 1S2, Qu\'ebec, Canada}

\begin{abstract}
%In a laser-plasma accelerator, although high quality electron beams are produced in the bubble/blow out regime, the shot to shot stability is not yet excellent with current laser parameters.
Using a laser plasma accelerator, experiments with a 80 TW and 30 fs laser pulse demonstrated quasi-monoenergetic electron spectra with maximum energy over 0.4 GeV. 
This is achieved using a supersonic He gas jet and a sharp density ramp generated by a high intensity laser crossing pre-pulse focused 3 ns before the main laser pulse. 
By adjusting this crossing pre-pulse position inside the gas jet, among the laser shots with electron injection more than 40\% can produce quasi-monoenergetic spectra.
This could become a relatively straight forward technique to control laser wakefield electron beams parameters. \end{abstract}

\maketitle

The production of tunable and low energy spread electron beams is needed for compact accelerator development, efficient injection of electron beams into an undulator insertion device, and production of tunable X-ray radiation via Thomson laser scattering. Electrons acceleration by laser wakefield has generated a lot of interest since the first quasi-monoenergetic electron beams generation demonstration  \cite{Nature2004Mangles, Nature2004Geddes, Nature2004Faure}. 
Laser wakefield acceleration is achieved when an intense femtosecond laser pulse is focused at relativistic intensities $\geq 10^{18}$ W/cm$^{2}$, onto a gas jet target. Interacting with the quasi-instantaneously created under-dense plasma, the laser pulse excites a wakefield in which electrons can be trapped and accelerated to high energies in short distances. Control of the electron spectrum is closely related to the injection and trapping of the electrons. In the mentioned experiments the electrons are self injected from the plasma: the intensity of the driving laser pulse is high enough to create a cavity region free from electrons where strong radial electric fields induce transverse and localized injection in the laser wakefield  \cite{PRE1998Bulanov}. In this regime, quasi-monoenergetic electron beams with maximum energies near 1 GeV have been reported in an experiment using a cm scale long capillary waveguide target \cite{NP2006Leemans}. With the current laser technology, the transverse injection threshold is reached due to the non linear evolution of the laser pulse during its propagation. Thanks to strong self focusing and self compression, it allows to produce the suitable conditions for transverse injection, resulting in a difficult control of the electron beam parameters. To overcome this problem, external injection is difficult as it requires injection of a femtosecond electron bunch with femtosecond timing due to the short plasma wavelength ($\lambda_p  \sim$ 10 $\mu$m). Thus, different techniques to obtain a better electron injection control have been proposed, such as ionization induced trapping, optical injection, and the use of a downward density ramp.

Electron trapping using a gas mixture has been recently demonstrated \cite{PRL2010Mcguffey,PRL2010Pak,PRL2010Clayton}. High Z gas for ionization induced trapping has also allowed the production of quasi-monoenergetic electron beams up to the 0.5 GeV range but with a small reported charge and a low energy tail due to continuous injection \cite{APL2012Mo}. These techniques could be a viable avenue to produce quasi-monoenergetic electron beams but further development is still needed. 
Optical injection requires the use of an additional laser beam to inject electrons \cite{PRLUmstadter96}. The most simple geometry makes use of two counter-crossing laser beams: one to drive the wakefield and the second to produce a standing wave that will pre-accelerate and inject electrons in the accelerating structure. High quality, stable quasi-monoenergetic electrons have been achieved using this method. The difficulty is that both $\mu$m spatial superposition and fs temporal synchronization are required for the two laser beams which is challenging \cite{Nature2006Faure,PRL2009Kotaki}. 
For the downward density ramp approach, two regimes have to be considered depending of $L_{grad}$, the electron density gradient. First, if $L_{grad} > \lambda_p$ a slowing down of the plasma wave velocity occurs that lowers the threshold for wave breaking of the wakefield and causes plasma background electrons trapping in a specific position of the density ramp  \cite{PRE1998Bulanov,PRST2003Tomassini}. Second, if $L_{grad} \le \lambda_p$ the plasma density transition produces a sudden increase of the plasma wavelength and causes the rephasing of a fraction of the plasma electrons into the accelerating phase of the plasma wave \cite{PRL2001Suk,JOSAB2004Suk}. Experimental implementation of a density modulation has been achieved using several techniques. A laser pre-pulse crossing the laser main beam propagation axis close to $90^\circ$ has been used to ionize and heat the gas locally to generate a plasma channel. At higher laser intensities the same result is obtained via Coulomb explosion. Using this technique $L_{grad}$ in various ranges have been achieved \cite{PRL2005Chien,PRE2004Kim,PoP2010Faure}. Geddes et al. used the slow downward density ramp produced at a gas jet exit to generate a density ramp \cite{PRL2008Geddes}. To obtain a sharp transition Schmid et al. used the shock front induced by a knife edge inserted in a gas jet \cite{PRST2010Schmid}.

\begin{figure*}
   \centering
   \includegraphics[width=0.9\textwidth]{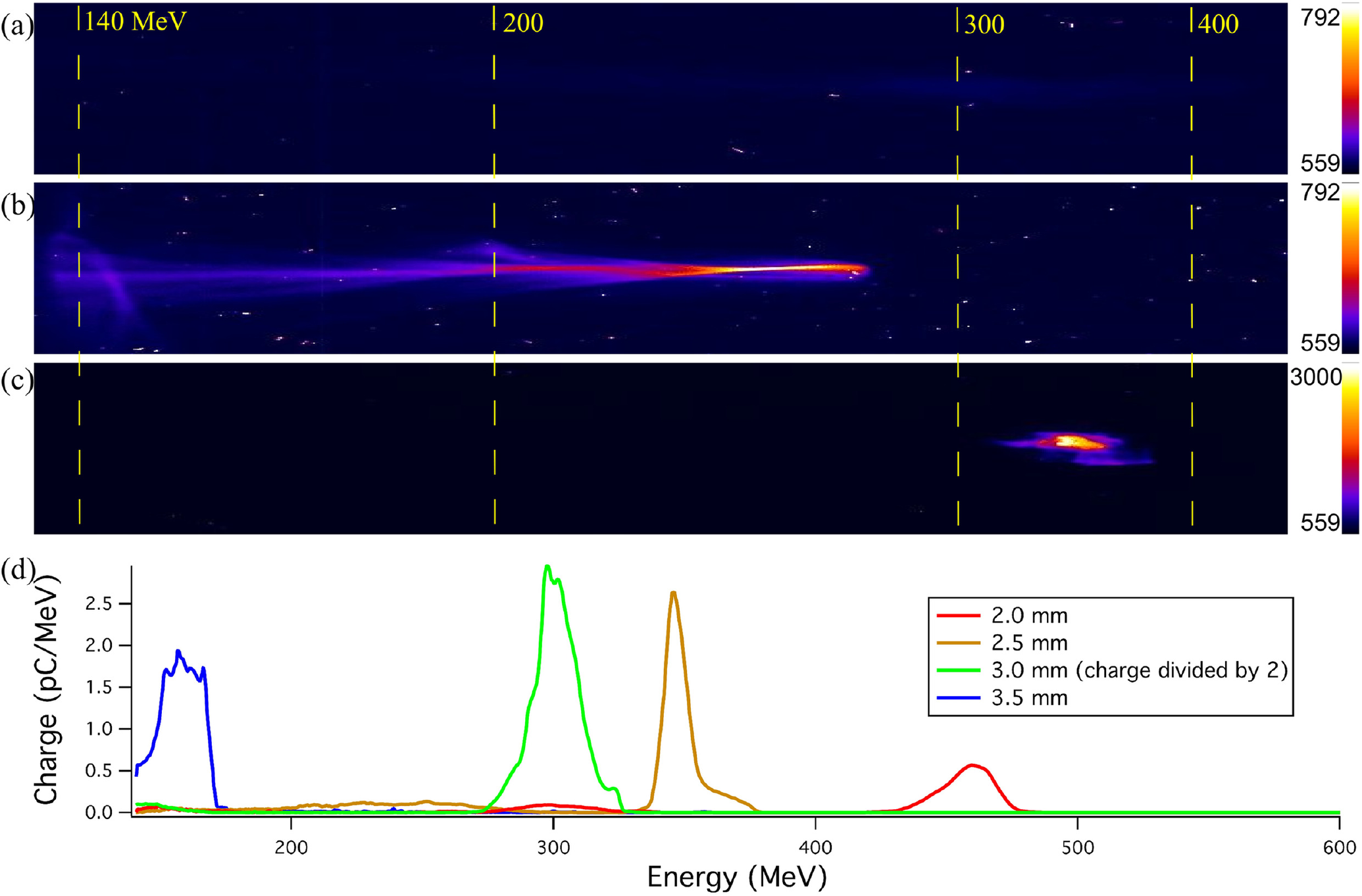}
   \caption{Electron spectra. (a) and (b): raw spectrum without laser crossing pre-pulse at electron density $5.2\times10^{18}$cm$^{-3}$ and $8.0\times10^{18}$cm$^{-3}$ respectively. (c): raw spectrum with laser crossing pre-pulse focused at $z =$ 3 mm and electron density $5.2\times10^{18}$cm$^{-3}$. (d): calibrated spectra with laser crossing pre-pulse focused at $z =$ 3.5, 3, 2.5 and 2 mm (respectively blue, green, brown, and red) and electron density $5.2\times10^{18}$cm$^{-3}$. Note that the charge has been divided by 2 for position $z=$ 3 mm to keep the charge scale. The intensity scale in counts is indicated on the left for pictures (a), (b), and (c). Note that the intensity scale is different on picture (c) from pictures (a) and (b).}
   \label{fig1}
\end{figure*} 

The few works using a sharp density ramp have been reported with limited laser power $\leq 10$ TW  \cite{PRL2005Chien,PRST2010Schmid}. Thus, it is of interest to study wakefield acceleration in a sharp density transition at higher laser powers. This has motivated the current experimental investigations: we report that a sharp 10 $\mu$m scale density transition produced by a high intensity laser crossing pre-pulse can be used to accelerate electrons with a probability of getting a quasi-monoenergetic beam higher than 40\% (among the laser shots with electron injection) and a mean energy up to 0.4 GeV range using 80 TW laser pulses. 

The experiment has been performed at the Advanced Laser Light Source (ALLS) facility at INRS-EMT using a 100 TW scale laser system. For our experimental conditions, the laser system produces 2.5 J of energy on target with a Full Width Half Maximum (FWHM) duration of 30 fs (80 TW) and linear polarization. The main laser pulse is focused onto a supersonic helium gas jet. In the focal plane, the FWHM spot size was 18 $\mu$m, with 50\% of the total energy contained within an area limited by the $1/e^2$ radius. This corresponds to a laser intensity of 1.2 $\times$ 10$^{19}$ W/cm$^2$ and a normalized vector potential amplitude $a_0 =2.4$. A fraction of the main laser pulse is taken to be used as a crossing pre-pulse to produce the density transition. This laser crossing  pre-pulse is focused onto the supersonic gas jet with a propagation axis at $90^\circ$ to the main laser pulse and a 3 ns delay. This time delay was chosen in order to obtain a steep electron density gradient and lead to enhanced electron injection \cite{PRL2005Chien,PoP2010Faure}. The laser crossing pre-pulse intensity at focus is $10^{18}$ W/cm$^2$. For these measurements, we used a 5 mm diameter supersonic helium gas jet producing a density profile well defined by a 4-mm-long electron density plateau. The laser crossing pre-pulse position of focus along the main pulse propagation axis is denoted z. Position $z = 0$ mm corresponds to where the main laser pulse enters the gas jet and $z = 5$ mm where it exits (the density plateau begins at 0.5 and ends at 4.5 mm). 
The electron beam produced in the interaction is measured using a spectrometer located inside the vacuum interaction chamber. It consists of two consecutive permanent dipoles magnets (1.1 T and 0.8 T respectively each over 10 cm long) that deflect electrons depending on their energy and a Lanex phosphor screen to convert a fraction of the electron energy into light imaged by a high dynamics cooled CCD camera. Calibration of the electron charge has been achieved by determining the CCD camera detection efficiency and published calibration data for an identical Lanex screen \cite{RSI2006Glinec}. The accuracy for the charge determination is estimated to be 18\%. Calibration of the electron energy is calculated by measuring the magnets field maps. The energy range lower limit is 120 MeV due to the maximum deflection angle inside the magnets. The spectrometer relative energy resolution is limited by the electron beam divergence. Assuming 10 mrad divergence, it is between 5 and 10\% for the interval 120-480 MeV.

Figure \ref{fig1} shows typical electron spectra. In Figs. \ref{fig1}(a) and \ref{fig1}(b), only the main laser pulse is used at two electrons densities: $5.2\times10^{18}$ cm$^{-3}$ and $8.0\times10^{18}$ cm$^{-3}$, respectively. The self injection threshold has been determined to be $6.0\times10^{18}$ cm$^{-3}$. For the lower density case (Fig. \ref{fig1}(a)), large energy spread peaked spectra with mean energy close to 300 MeV and few pC total charge are observed for 30\% of the laser shots (otherwise no signal is detected). The beam divergence measured at $1/e$ is 7 mrad. For the higher density case (Fig. \ref{fig1}(b)), we observe electron spectra with large energy spread on all the laser shots and 134 pC average total charge. The beam divergence is 8 mrad. To test the injection into a sharp density gradient we measured the electron spectra at the lower density (Fig. \ref{fig1}(a)), where no significant self injection occurs, as a function of $z$. $L_{grad}$ is estimated with a Thomson self-scattering diagnostic of the laser pulse during its propagation. The light emitted at $90^\circ$ from the laser polarization is imaged with a CCD camera. The laser crossing pre-pulse propagation produces an electron density depression and a spike on its edge. On the optical diagnostic, this corresponds to two light intensity maximum due to the scattering on the density spike at the edge of the plasma channel. We find the two intensity maxima to be separated by 120 $\mu$m and estimate $L_{grad} \sim 25$ $\mu$m from the intensity profile ($\lambda_p = 14.6$ $\mu$m at $5.2\times10^{18}$ cm$^{-3}$ electron density). Thomson scattering is not a direct diagnostic of the plasma density gradient but has been found in good agreement with the results obtained from interferometric techniques \cite{PRL2005Chien,PoP2010Faure}. We illustrate the effect of the laser crossing pre-pulse over the injection process with Fig. \ref{fig1}(c) where the spectrum corresponds to the best conditions in term of stability and charge for quasi monoenergetic beam production at $z=3$ mm. On this particular shot the mean energy is $E_m = 335$ MeV, the energy spread is $\Delta E/E = 8\%$ (calculated with the energy spread at $1/e$ of the mean energy), the beam divergence is 5 mrad, and the charge contained into the $1/e^2$ area around $E_m$ (energy and divergence axis) is 335 pC . In Fig. \ref{fig1}(d) shots obtained using the main laser pulse and the crossing pre-pulse are shown respectively for $z =$ 3.5, 3, 2.5 and 2 mm (blue, green, brown and red lines). Only electron beam with spectra exhibiting monoenergetic features are shown. Note that the charge has been divided by 2 for position $z=3$ mm to keep the charge scale. We observe that the energy can be tuned depending of the $z$ position. To illustrate the capability of this technique to generate high energy quasi-monoenergetic electron beams we show on this picture (red line) a laser shot (20\% of shots for $z =$ 2 mm) with mean energy close to 0.5 GeV. For this spectrum, the mean energy is $E_m = 470$ MeV, the energy spread is $\Delta E/E = 6\%$, the beam divergence is 10 mrad and the charge contained around $E_m$ is 40 pC. 
 
\begin{figure}
   \centering
   \includegraphics[width=8.5cm]{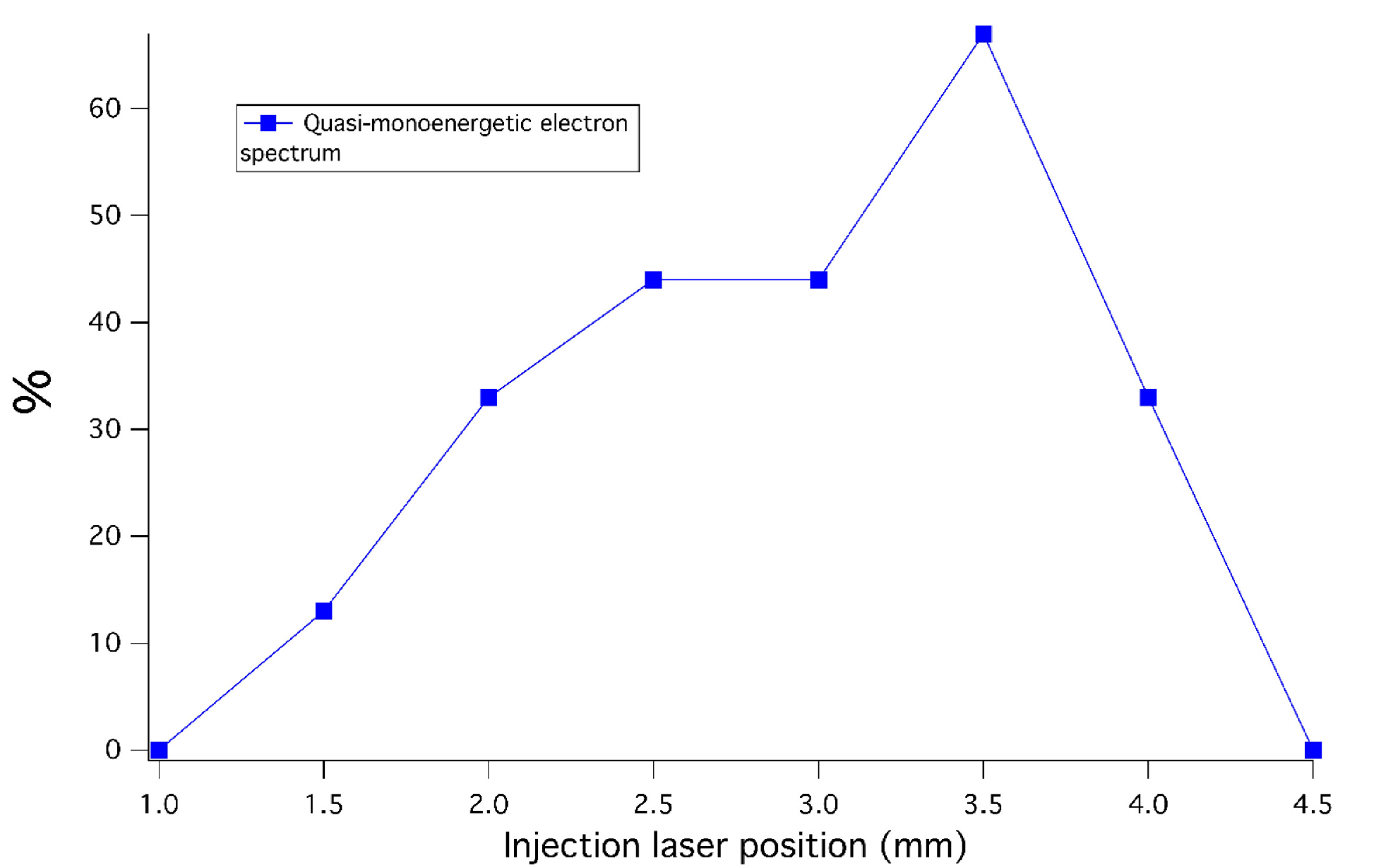}
   \caption{Probability to get quasi-monoenergetic spectra as a function of the crossing pre-pulse $z$ position.}
   \label{fig2}
\end{figure} 

\begin{figure}
   \centering
   \includegraphics[width=8.5cm]{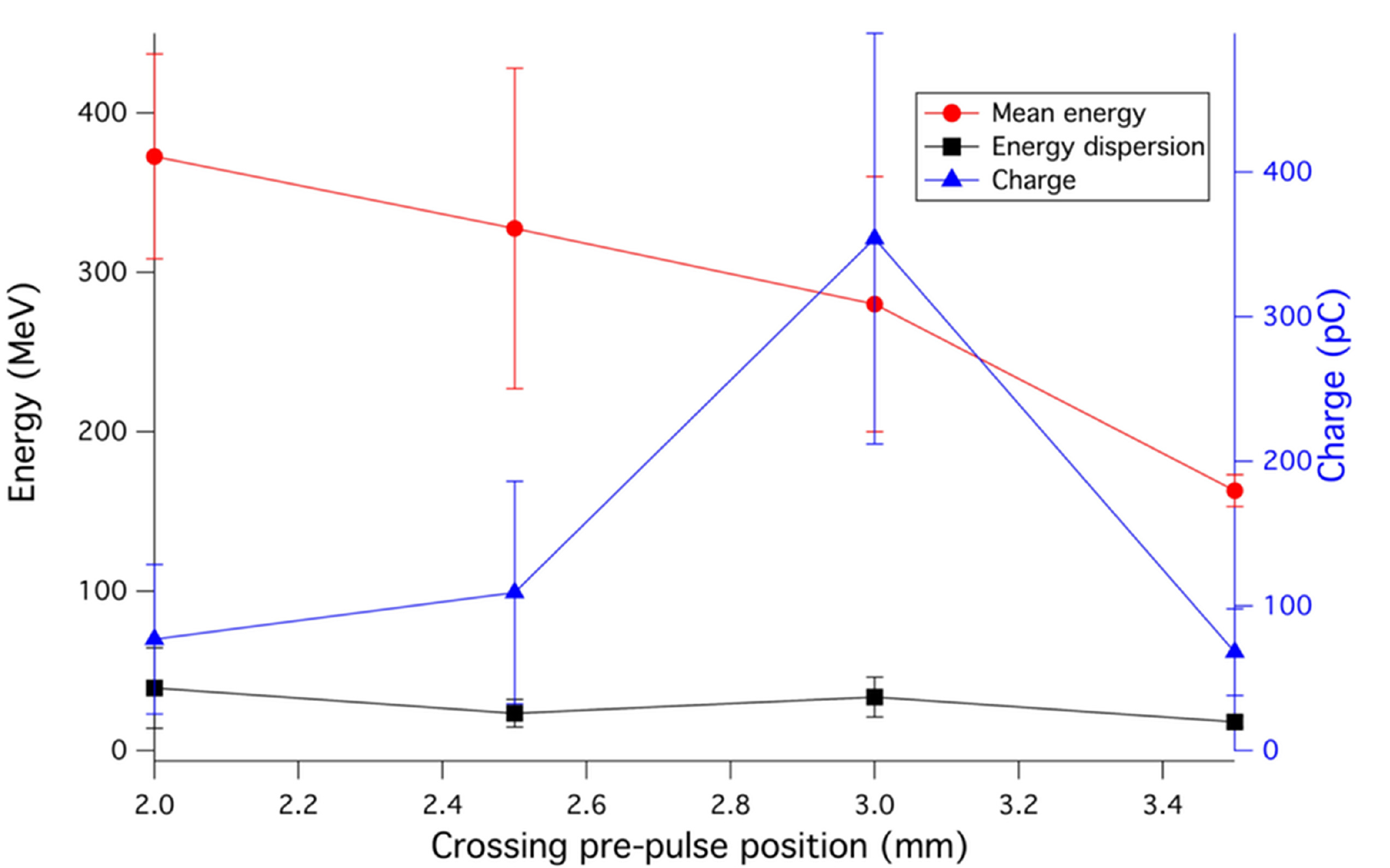}
   \caption{Evolution of the mean energy (red circle), charge (blue triangle), and energy spread (black square) of electron spectra as a function of the crossing pre-pulse $z$ position. Each point show the average over all the quasi-monoenergetic electrons beams.}
   \label{fig3}
\end{figure}

The effect of the laser crossing pre-pulse to inject electrons is clearly observed but shot to shot variations of the electron spectra features are present. 
The probability to inject the electrons (with or without monoenergetic features) is $\geq90\%$ for the interval $z=$ 2-3 mm and $\geq50\%$ at 3.5 mm.
There is no electron signal before $z =$ 1 mm because it corresponds to the propagation phase of the laser.
Starting from $z=4$ mm and higher, the electron charge decreases strongly as the main laser reaches the end of the gas jet. 
Figure \ref{fig2} shows, among the laser shots with electron injection, the probability to produce a quasi-monoenergetic spectrum as a function of the laser crossing pre-pulse position $z$. The maximum probability to generate such spectrum is greater than 40\% for the interval $z=$ 2.5-3.5 mm.
It is interesting to notice on Fig.~\ref{fig2} that the probability to generate a quasi-monoenergetic electron beam increases with its position along the interaction distance, reaches a maximum and decreases. This is a typical laser beam propagation behavior, which self focuses after a given length before being refracted.
Figure~\ref{fig3} shows, for laser shots generating quasi-monoenergetic electron beams, the mean energy $E_m$ (red circle), the charge around the mean energy (blue triangle), and the energy spread $\Delta E$ (black square) as a function of the laser crossing pre-pulse $z$ position. The error bars correspond to the standard deviation for the measured laser shots . Only the position interval that corresponds to the highest probability to generate quasi-monoenergetic electron beams combined with a high charge (over 50 pC) is shown. In this position range we observe that the mean energy can be tuned between 150 and 370 MeV.  
As expected the electron energy increases almost linearly with the acceleration length which is controlled by changing the crossing pre-pulse position that creates the density gradient. 

An optimum compromise is found at $z=3$ mm where an average charge greater than 300 pC is obtained combined with the following average parameters: mean energy of 280 MeV, energy spread $\Delta E/E = 11 \%$, and divergence 6 mrad. 
At this position, the shot to shot standard deviation is 28\% for the mean energy and 40\% for the electrons charge.

In conclusion, we demonstrated using  a sharp density transition the generation of quasi-monoenergetic electron spectra with maximum energy over 0.4 GeV and an average charge close to 100 pC. 
By adjusting the crossing pre-pulse position inside the gas jet, among the laser shots with electron injection more than 40\% can produce quasi-monoenergetic spectra.
This technique, with further optimization, could become a relatively straight forward method to control laser wakefield electron beams. 

\bigskip
We thanks ALLS technical team for their support. 
The ALLS facility was funded by the Canadian Foundation for
Innovation (CFI). This work is funded by NSERC and the Canada Research Chair
program.
We acknowledge the Agence Nationale pour la Recherche, through the COKER project ANR-06-BLAN-0123-01, 
and the European Research Council through the PARIS ERC project (Contract No. 226424) for their financial supports.

\end{document}